\documentstyle[amstex,aps,12pt]{revtex}

\def\EE{{\hbox{I\kern-.2em\hbox{E}}}}

\input epsf
\begin{document}
\begin{titlepage}
\thispagestyle{empty}
\title{ VERIFICATION OF A LOCALIZATION CRITERION \vskip 0.1cm 
FOR SEVERAL DISORDERED MEDIA}
\vskip 0.5cm
\author{C. Ordenovic \and
G. Berginc\thanks{Thomson CSF-Optronique, rue Guynemer, BP 55, 78283 
Guyancourt cedex, France.} and C. Bourrely, \vskip 0.2cm
Centre de Physique Th\'eorique\thanks{Unit\'e Propre de Recherche 7061} 
- CNRS - Luminy Case 907, 
\vskip 0.1cm 
13288 Marseille Cedex 09, France}
\maketitle
\vskip 1.5cm
\begin{center}
\bf{Abstract}
\end{center}
\begin{abstract}
We analytically compute a localization criterion in double
scattering approximation for a set of dielectric spheres or perfectly
conducting disks uniformly distributed in a spatial volume which
can be either spherical or layered. For every disordered medium, we 
numerically investigate a localization criterion, and examine the 
influence of the system parameters on the wavelength localization domains.
\end{abstract}

\vskip 2.5cm
\normalsize
\noindent Key words : localization, disordered media, electromagnetic
scattering

\noindent Number of figures: 20

\noindent May 1999

\noindent CPT-99/P.3815

\noindent anonymous ftp: ftp.cpt.univ-mrs.fr\\
\noindent Web address: www.cpt.univ-mrs.fr

\end{titlepage}

\section{INTRODUCTION}

Several works of Anderson on localization \cite{Anderson} have mostly been 
related
to electrons transport in solids with random impureties.
Wolfle \cite{wolfe} has proposed a diagrammatical treatment for the electronic
localization in a bidimensional disordered medium. This formalism was extended
to the electromagnetic waves in a volume of disordered medium \cite{Arya},
\cite{Sheng}, where the impureties are modeled by dielectric spherical
scatterers, and the role played by the backscattering mechanism in the energy
localization was established. The physical basis of the localization is a
consequence of the vanishing of diffusion coefficient of the
electromagnetic wave energy, this condition was then called a localization 
criterion.

The purpose of this paper is to analytically determine several conditions on 
the parameters of some disordered systems (namely,  
the number of scatterers, their size, permittivity, geometry, the width 
of the medium) satisfying a localization criterion of the electromagnetic 
energy, and to show their influence on the wavelength localization
domains \cite{ordenov}. We give the analytic expressions of the
localization criterion for a spherical or layered medium made of dielectric
spheres or perfectly conducting disks.
We numerically determine the sets of parameters offering ranges in 
wavelength where the localization criterion is satisfied. 
More precisely, we compute the evolution of the range when we modify 
every parameter, and we show that the location and the bandwitdh of these 
ranges may be controlled by varying some characteristic parameters of 
the disordered system.

This paper is organized as follows. After the introduction, we recall in
Section 2 the mathematical model used to describe the behaviour of the
electromagnetic wave in a disordered medium, and the expression of a
localization criterion. Then in Section 3, we derive analytically a
localization criterion for several disordered media. In Section 4, we
numerically compute the localization criterion for these media in order to
determine localization domains, and we examine the influence of the system
parameters on these localization domains. Computations  with a multidipolar
formulation are used for comparison with the theoretical model. 
Finally, Section 5 is devoted to the conclusions.

\section{THE MATHEMATICAL MODEL}

\begin{figure}[b]
\epsfxsize=10cm
\centerline{\epsfbox{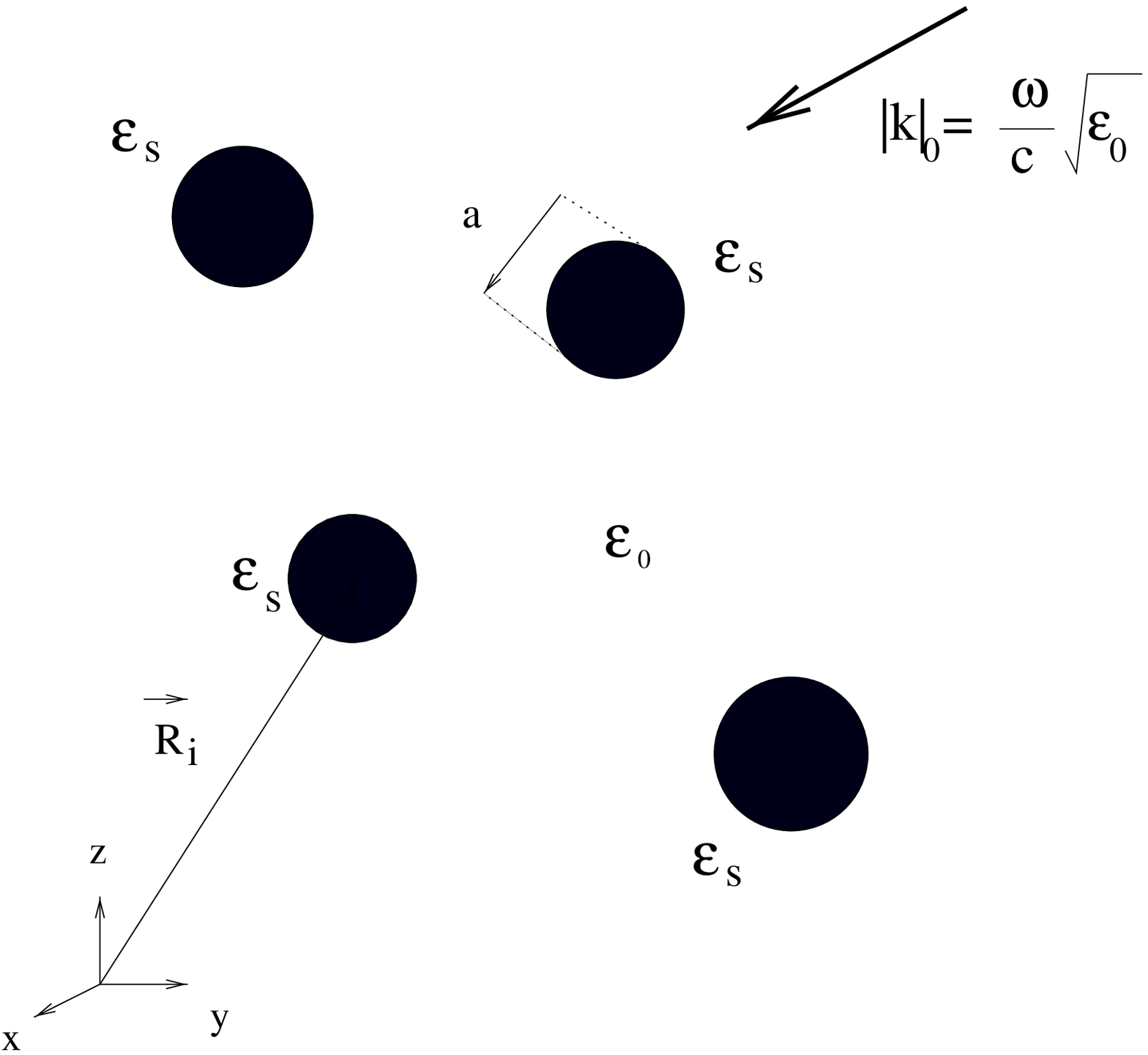}}
\caption{Scheme of the disordered medium.}
\label{fi:fig1}
\end{figure}

We consider $N$ identical scatterers of relative permittivity
$\Bar{\Bar{\epsilon_{s}}}(\omega)$ distributed in a spatial volume $V$. The
location of scatterers is given by the vector $\overrightarrow{ R_{p}}, p
\in[1,N]$. The background medium has a relative permittivity
$\epsilon_{0}$.
An incident electromagnetic plane wave of pulsation $\omega$, and
wavevector $k_{0}= \displaystyle{\frac{\omega}{c}}\sqrt{\epsilon_{0}}$, 
interacts with the set
of scatterers  (figure \ref{fi:fig1}).

\noindent
The electric field $\overrightarrow{ E}(\overrightarrow{ r},\omega)$
 at location $\overrightarrow{ r}$ satisfies the Helmholtz equation
\begin{align}
\Delta \overrightarrow{ E}(\overrightarrow{ r},\omega) + k^2
\Bar{\Bar{\epsilon}}(\overrightarrow{ r},\omega).\overrightarrow{
E}(\overrightarrow{ r},\omega) = 0,
\end{align}
\noindent
where $k=\displaystyle{\frac{\omega}{c}}$ is the wavevector in the vacuum,
$\Bar{\Bar{\epsilon}}(\overrightarrow{ r},\omega)$ is the second rank tensor
associated with the relative permittivity of the medium, 
where the components are given by
\begin{align}
\epsilon_{ij}(\overrightarrow{ r},\omega)  =  \epsilon_{0}\delta_{ij} +
\sum_{p=1}^{N} \big(\epsilon_{s ij}(\omega) - \epsilon_{0}\delta_{ij}\big)
\chi_{\overrightarrow{ R_{p}}}(\overrightarrow{ r}),
\end{align}
$\chi_{\overrightarrow{ R_{p}}}(\overrightarrow{ r})$ is the characteristic
function of the $p^{th}$ scatterer
\begin{align}
\chi_{\overrightarrow{ R_{p}}}(\overrightarrow{ r}) = \begin{cases}
                         1 \quad \text{if $\overrightarrow{ r}$ inside the
scatterer,}\\
                         0 \quad \text{outside}.
                             \end{cases}
\end{align}
\noindent
We assume the scatterers homogeneous and isotropic, thus
\begin{align}
\epsilon_{ij}(\overrightarrow{ r},\omega)  =  \bigg(\epsilon_{0} +
\sum_{i=1}^{N} \big(\epsilon_{s}(\omega) - \epsilon_{0}) \chi_{\overrightarrow{
R_{i}}}(\overrightarrow{ r})\bigg) \delta_{ij},
\end{align}
\noindent
so we can treat the relative permittivity of the medium as a function of
$\overrightarrow{ r}$ and $\omega$.

\noindent
Then, we write the Helmholtz equation in a perturbative form
\begin{align}
\label{perturb}
\Delta \overrightarrow{ E}(\overrightarrow{ r},\omega) + \big[k_{0}^2 - \delta
V(\overrightarrow{ r})\big] \overrightarrow{ E}(\overrightarrow{
r},\omega) = 0,
\end{align}
\noindent
where $\delta V(\overrightarrow{ r}) =
k^2(\epsilon_{0}-\epsilon_{s})\sum_{i=1}^{N}\chi_{\overrightarrow{
R_{i}}}(\overrightarrow{ r})$ is the potential function describing the
perturbation of the incident wavevector.

\subsection{The Green's formalism}

\noindent
In order to solve the Helmholtz Eq.~(\ref{perturb}), we first introduce
the dyadic Green function 
$g_{ij}(\overrightarrow{ r},\overrightarrow{r'},\omega)$ 
\cite{Tai} which determines the wave at location $\overrightarrow{r}$ 
generated by a pointlike current source located in $\overrightarrow{ r'}$.
Its components satisfy the Helmholtz equation
\begin{align}
\Delta g_{ij}(\overrightarrow{ r},\overrightarrow{ r'},\omega) + \big[k_{0}^2 -
\delta V(\overrightarrow{ r})\big] g_{ij}(\overrightarrow{
r},\overrightarrow{ r'},\omega)
= \delta_{ij}\delta(\overrightarrow{ r}-\overrightarrow{ r'}).
\end{align}
\noindent
The interest of introducing a dyadic Green function is to provide a generic 
formulation for the electromagnetic field behaviour.

\noindent
This equation can be written in an integral form
\begin{align}
g_{ij}(\overrightarrow{r},\overrightarrow{r}',\omega) =
g^{0}_{ij}(\overrightarrow{r} - \overrightarrow{r}',\omega)
+ \sum_{k=1}^{3}\int g^{0}_{ik}(\overrightarrow{r} -
\overrightarrow{r_{1}},\omega) \delta V(\overrightarrow{r_{1}})
g_{kj}(\overrightarrow{r_{1}},\overrightarrow{r}',\omega)
d\overrightarrow{r_{1}}, \label{green0}
\end{align}
\noindent
where $g^{0}_{ij}(\overrightarrow{r}-\overrightarrow{r'},\omega)$ is the dyadic
Green function of the electromagnetic wave in the nonperturbated medium of
relative permittivity $\epsilon_{0}$, which satisfies the equation
\begin{align}
g^{0}_{ij}(\overrightarrow{r} - \overrightarrow{r}') = \bigg( \delta_{ij} +
\partial_{i} \partial_{j} \bigg) g^{0}(\overrightarrow{r} -
\overrightarrow{r}'),
\end{align}
\noindent
and $g^{0}(\overrightarrow{r} - \overrightarrow{r}')$ is the scalar Green
function of the wave
\begin{align}
\label{green_libre}
g^{0}(\overrightarrow{r} - \overrightarrow{r}',\omega) = - \frac{1}{4\pi}
\frac{e^{i\ k_{0}|\overrightarrow{r} -
\overrightarrow{r}'|}}{|\overrightarrow{r} - \overrightarrow{r}'|}.
\end{align}
\noindent
We introduce the Green operators ${\mathcal G}$ and ${\mathcal G}^{0}$
\begin{align}
{\mathcal G}^{0} \quad &: \quad \overrightarrow{\Psi}(r) \rightarrow\int
\Bar{\Bar{g^{0}}}(r-r')\overrightarrow{\Psi}(r') dr',\\
{\mathcal G} \quad &: \quad \overrightarrow{\Psi}(r) \rightarrow\int
\Bar{\Bar{g}}(r,r')\overrightarrow{\Psi}(r') dr'.
\end{align}
The dyadic Green functions are defined by the kernels of the operators. The
operator $\delta V$ in Eq. (\ref{green0})
is defined as the multiplicative operator,
and the direct product between two successive operators is written 
as a point.

\noindent
Using this formalism, we can rewrite the integral equation like
\begin{align}
\label{integrale}
{\mathcal G} = {\mathcal G}^{0} + {\mathcal G}^{0}.\delta V .{\mathcal G}.
\end{align}

\subsection{First moment of ${\mathcal G}$ - Self energy operator}

\subsubsection{Model of disordered medium}

The calculation of the kernel of ${\mathcal G}$ using a perturbative expansion
of (\ref{integrale}) requires the knowledge of the $3N$ space coordinates
relative to the location of scatterers. Due to their complexity, we introduce 
a statistical approach of the problem. We call
$\rho_{N}(\overrightarrow{R_{1}}...\overrightarrow{R_{N}})$ the probability
density of location of the $N$ scatterers in the medium \cite{Frisch}, and we
introduce the mathematical expection value or mean $\EE$ defined by
\begin{align}
\label{eesper}
\EE\big({\mathcal G}\big) : \overrightarrow{\psi}(\overrightarrow{r})
\rightarrow \int_{V} {g}_{ij}(\overrightarrow{r}-\overrightarrow{r}',\omega)
\rho_{N}(\overrightarrow{R_{1}}...\overrightarrow{R_{N}})
\Pi_{p=1}^{N}d\overrightarrow{R}_{p}\psi(\overrightarrow{r}')d
\overrightarrow{r}',
\end{align}
where we average over the whole possible configurations of the disordered
medium.

\subsubsection{First moment of ${\mathcal G}$}

From Eq.~(\ref{integrale}), (\ref{eesper}) we obtain
\begin{align}
\label{firstmom}
\EE\big({\mathcal G}\big) = {\mathcal G}^{0} + {\mathcal G}^{0} 
\EE\big(\delta V . {\mathcal G}\big).
\end{align}
By introducing the self-energy operator $\Sigma$ \cite{Arya}, defined by
\begin{align}
\label{sig1}
\Sigma = \EE\big(\delta V . {\mathcal G}\big) \big[\EE\big({\mathcal G})\big]^{-1},
\end{align}
we can rewrite (\ref{firstmom}) like a Dyson equation
\begin{align}
\label{Dyson}
\EE\big({\mathcal G}\big) = {\mathcal G}^{0} + {\mathcal G}^{0} \Sigma
\EE\big({\mathcal G}\big).
\end{align}
\noindent
The definition (\ref{sig1}) of $\Sigma$  is not in a convenient form
to obtain a solution, because
$\EE\big({\mathcal G}\big)$ is the unknown of the equation. So, in order to
calculate $\Sigma$, we prefer to link it with the multiple scattering formalism
\cite{Arya}. If ${\mathcal T}$ is the scattering operator of the medium,
${\mathcal T}$ is defined by
\begin{align}
\label{deftau}
{\mathcal G} = {\mathcal G}^{0}+{\mathcal G}^{0}.{\mathcal T}.
{\mathcal G}^{0},
\end{align}
and the expectation value $\EE({\mathcal T})$ of the scattering operator is
given by
\begin{align}
\label{expece}
\EE({\mathcal T}) = {\mathcal G}^{0}+{\mathcal G}^{0}.\EE({\mathcal T}).
{\mathcal G}^{0}~.
\end{align}
We have the following relation between $\Sigma$ and $\EE({\mathcal T})$,
\begin{align}
\label{devsigma}
\Sigma = \EE({\mathcal T})\bigg[ 1 + {\mathcal G}^{0}.\EE({\mathcal T})\bigg]^{-1}
= \EE({\mathcal T}) - \EE({\mathcal T}). {\mathcal G}^{0}. \EE({\mathcal T}) + ...,
\end{align}
\noindent
where ${\mathcal T}$ can be expressed using a multiple scattering expansion in
terms of the scattering operator by a single scatterer $t_{i}$ at location
$\overrightarrow{R}_{i}$,
\begin{align}
\label{multidif}
{\mathcal T} = \sum_{i=1}^{N}t_{i} + \sum_{i=1}^{N} 
{t}_{i}{\mathcal G}^{0}\cdot\sum_{j=1 , j\neq i}^{N} t_{j}
+ \sum_{i=1}^{N} t_{i}{\mathcal G}^{0}\cdot\sum_{j=1 , j\neq i}^{N} 
t_{j}{\mathcal G}^{0}\cdot\sum_{k=1 , k \neq j}^{N} t_{k} + ...
\end{align}
The interest of the last expression is to express $\Sigma$ by the mean of a 
scattering theory instead of calculating it from a perturbative expansion.

\subsubsection{Macroscopic homogenization of the disordered medium - Scalar
formulation}

\noindent
We use the hypothesis of macroscopic homogenization of the disordered medium
in order to calculate the kernel of $\EE\big({\mathcal G}\big)$, 
i.e., the dyadic Green function of the averaged wave. 
We suppose that after averaging, the medium behaves like a homogeneous
one, which implies an invariance under translation. In consequence, we can 
first write the Dyson equation (\ref{Dyson}) like a convolution equation
\begin{align}
\label{conv}
\EE\big({\mathcal G}\big) = {\mathcal G}^{0} +
{\mathcal G}^{0}*\Sigma*\EE\big({\mathcal G}\big).
\end{align}
\noindent
The other consequence of the macroscopic homogenization hypothesis is the
possibility to write a scalar form of the previous equation. Indeed, the
averaged medium being homogeneous, the kernel of $\EE\big({\mathcal G}\big)$
may be written as
\begin{align}
\EE\big(g(\overrightarrow{r}-\overrightarrow{r}')\big)_{ij} =
\bigg(\delta_{ij} +
\partial_{i}\partial_{j}\bigg)
\EE\big(g(\overrightarrow{r}-\overrightarrow{r}')\big),
\end{align}
\noindent
where $\EE\big(g(\overrightarrow{r}-\overrightarrow{r}')\big)$ 
is the first moment of the scalar
Green function of the electromagnetic wave. The same consideration holds
for the kernel of the self-energy operator. Extracting the scalar part of
Eq.~(\ref{conv}), we obtain the scalar expression of the Green function
in Fourier space
\begin{align}
\EE\big(G(\overrightarrow{k},\omega)\big) = \frac{1}{k^2-k_{0}^2 -
\Sigma(\overrightarrow{k},\omega)},
\end{align}
\noindent
where we notice that the invariance by translation of the averaged medium
implies that the first moment of the Green function and the self energy 
function in Fourier space are only $\overrightarrow{k}$ dependent.
$\EE(G)$ is the scalar Green function of the homogenized medium.

The argument of Sheng \cite{Sheng} is to treat the averaged medium like an
effective medium, where the characteristic length of the averaged 
inhomogenities is small with respect to the wavelength, 
and therefore, they are not resolved by the wave, which explains that the 
averaged medium may be considered like a homogeneous one. 
In that case, the self-energy function $\Sigma$ describing the local 
microstructures of the medium, is weakly dependent on the variable
$\overrightarrow{k}$, so it is treated like a pulsation dependent 
function $\Sigma(\omega)$.

If this condition is satisfied, the first moment of the Green function
physically represents a propagative wave of effective wavevector 
${\tilde k}_{e}^{2} = k_{0}^2 + \Sigma(\omega)$, 
from which we deduce the wavevector of the wave
\begin{align}
\label{ke}
k_{e} = \mbox{Re}\big({\tilde k}_{e}\big) =
\sqrt{\frac{k_{0}^2+\Gamma}{2}}\bigg[1
+\sqrt{1+\big(\frac{\gamma}{k_{0}^2+\Gamma}\big)^2}\bigg]^{\frac{1}{2}},
\end{align}
\noindent
where $\Gamma = \mbox{Re}\big(\Sigma(\omega)\big)$,
$\gamma=\mbox{Im}\big(\Sigma(\omega)\big)$,
and the scattering length $l$ \cite{Arya} which is the inverse of twice 
the damping rate $\beta$ of the modulus of the averaged wave
\begin{align}
\label{l}
l=\frac{1}{2\beta}= \frac{k_{e}}{\gamma}.
\end{align}
\noindent
The decreasing factor in the expression of the first moment of the Green
function shows the fact that during the path in the averaged medium, the
wave loses its phase coherence over a characteristic length given by $l$, 
as a consequence of the successive scatterings. Over a distance of several
scattering lengths $l$, the first moment of the Green function becomes
negligible. In order to describe the behaviour of the electromagnetic wave
for large distances, we need to investigate the second moment of the dyadic
Green function.

\subsection{The second moment of ${\mathcal G}$ - Localization of the energy}

The averaged energy operator $\mathcal{P}$ is given by the averaged tensorial
product of the two Green operators ${\mathcal G}$ and ${\mathcal G}^{*}$, where
their kernels are respectively the dyadic Green function of the electromagnetic
wave at pulsation 
$\omega_{+} = \omega + \displaystyle{\frac{\delta \omega}{2}}$ and the
conjugate complex dyadic Green function  at pulsation 
$\omega_{-} = \omega - \displaystyle{\frac{\delta \omega}{2}}$
\begin{align}
\EE\big(\mathcal{P}\big) = \EE\big({\mathcal G} \otimes {\mathcal G}^{*}\big),
\end{align}
where $\omega$ is the central pulsation of the two dyadic Green functions, and
$\delta \omega$ is their pulsation difference. We notice that 
$\delta \omega$ is the conjugated variable of the time traject $t$ 
of the electromagnetic wave in the medium. 
The behaviour for $t \rightarrow +\infty$ of the averaged energy will
be given by the  behaviour of the kernel of
$\EE\big(\mathcal{P}\big)$ when $\delta \omega \rightarrow 0$.

\noindent
The outer product $\otimes$ applied on the components of the dyadic Green 
functions is given by
\begin{align}
\big(g \otimes g^{*}\big)
_{ijkl}(\overrightarrow{r},\overrightarrow{r}',\omega,\delta \omega) =
g_{ij}(\overrightarrow{r},\overrightarrow{r}',\omega_{+})
g^{*}_{kl}(\overrightarrow{r},\overrightarrow{r}',\omega_{-}).
\end{align}

Using the macroscopic homogenization hypothesis, we can write the scalar form
of the kernel of $\EE\big(\mathcal{P}\big)$, which is defined as the function
 $\EE\big(P(\overrightarrow{ r},\overrightarrow{ r'},\omega,\delta\omega)\big)$,
and is related to the average energy density function $L$ by its spatial 
Fourier transform
\begin{align}
\EE\big(P(\overrightarrow{ r},\overrightarrow{ r'},\omega,\delta\omega)\big) =
\frac{1}{2\pi}
\int L(\overrightarrow{ q},\omega,\delta \omega) e^{i \overrightarrow{
q}.(\overrightarrow{ r}-\overrightarrow{ r'})} d\overrightarrow{ q}.
\end{align}
Let us remark that the $q$ variable in the function $L$ is conjugated
with the wave traject $|\overrightarrow{r} - \overrightarrow{r}'|$. The
behaviour of the average energy at large distance will be given by the
behaviour of $L$ when $q \rightarrow 0$.

Following the frameworks of Arya\cite{Arya} and Sheng \cite{Sheng}, at large
distances ($q \rightarrow 0$) and large time
($\delta \omega \rightarrow 0$), $L$ is governed by a diffusion-like equation
\begin{align}
-i \delta \omega L(\overrightarrow{ q},\omega,\delta \omega) = q^2
D(\overrightarrow{ q},
\omega,\delta \omega) L(\overrightarrow{ q},\omega,\delta \omega),
\end{align}
where
\begin{align}
\label{dif}
D(\overrightarrow{ q},\omega,\delta \omega) \!\!= D^{B}(\omega) \bigg[1 -
\frac{\pi}{3 k_{e}^2 l ^2} \int_{Q=0}^{Q \sim \frac{1}{l}}
{Q^2 \over {i\delta \omega \over D(\bf q,\omega,\delta \omega)}
+ Q^2} d Q\bigg],
\end{align}
\noindent
is the general energy diffusion coefficient of the wave, and
\begin{align}
D^{B}(\omega) = \frac{\omega l}{3 k_{e}},
\end{align}
is the Boltzmann diffusion coefficient. The corrective term involved in the
general expression of the diffusion coefficient is a consequence of the
backscattering effect of the electromagnetic wave in the disordered medium.

\noindent
The energy of the electromagnetic wave is said to be localized when the general
diffusion coefficient (\ref{dif}) vanishes, i.e. the corrective term
generated by the backscattering compensates the Boltzmann diffusion
coefficient, which gives us the condition
\begin{align}
k_{e}.l \leq \sqrt{\frac{3}{\pi}},
\end{align}
\noindent
$k_{e}$ and $l$ are given by the relations (\ref{ke}) and (\ref{l}).
This condition is called the {\it localization criterion} of 
the electromagnetic wave energy.

\section{Calculation of the localization criterion \\ for several media}

\noindent
We propose to derive an analytic expression of the localization criterion 
for an electromagnetic wave of wavelength 
$\lambda_{0}=\displaystyle{\frac{2\pi}{k_{0}}}$ interacting with the two
following sets of media.

\begin{figure}
\epsfxsize=5cm
\centerline{\epsfbox{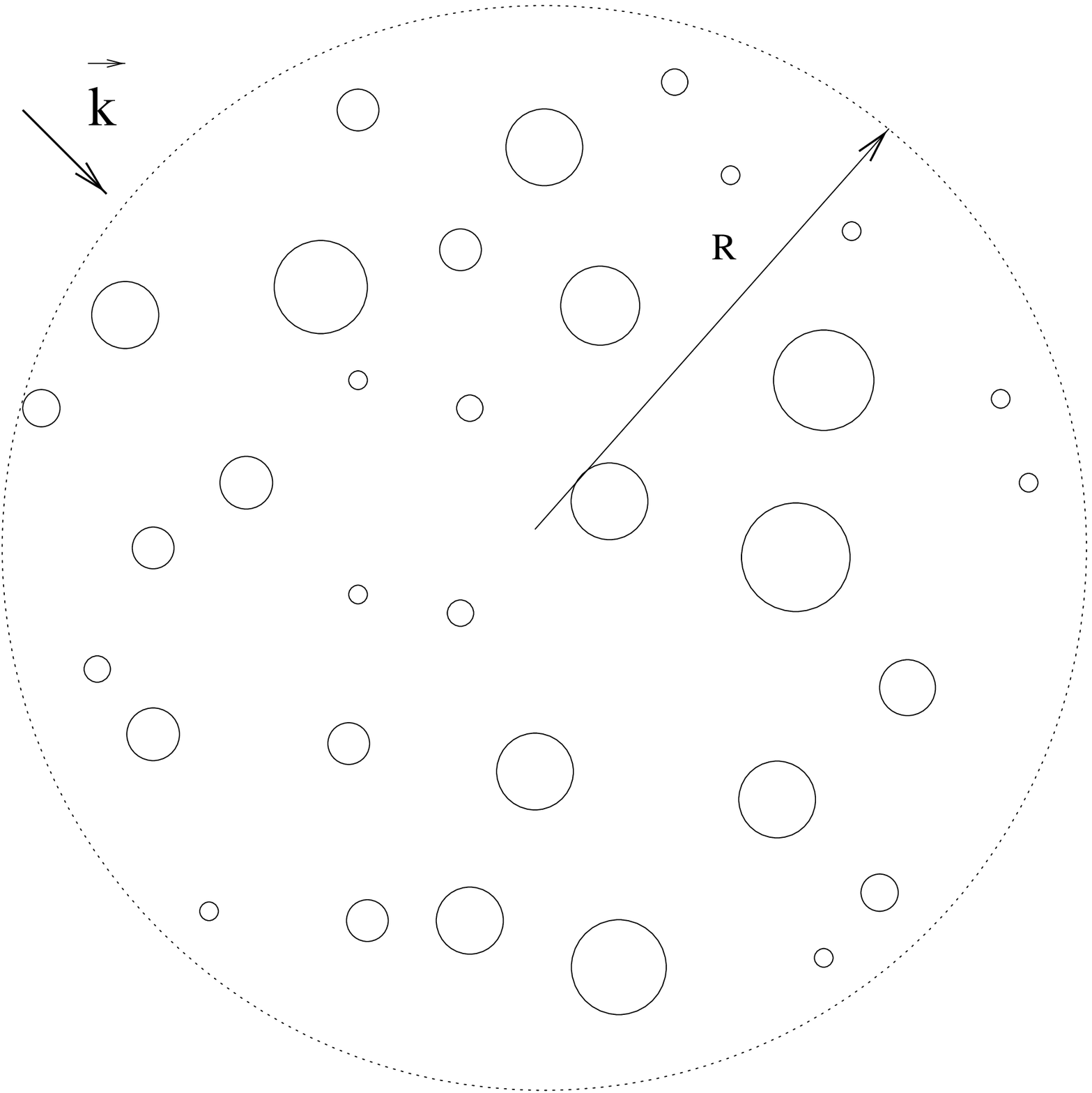}
\hskip 0.5cm 
\epsfxsize=5cm
\epsfbox{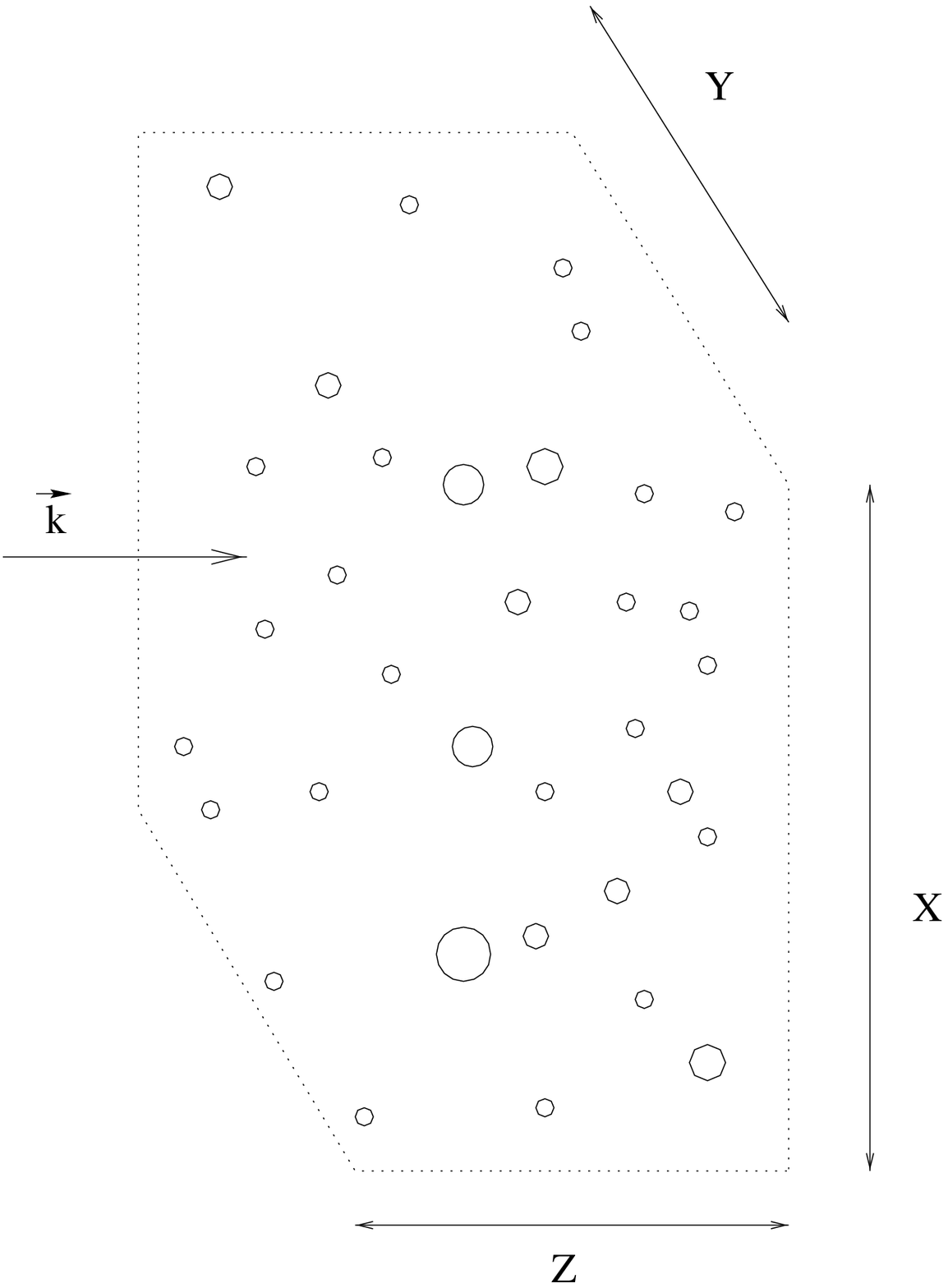}}
\vskip 2cm
\caption{Scheme for $N$ spheres uniformly distributed in a spherical
volume or in a layered volume of finite thickness.}
\label{fi:milieu_spheres}
\end{figure}

- A set of $N$ identical dielectric spheres of radius $a$ and relative
permittivity $\epsilon_{s}$, uniformly distributed in a spherical volume $V$ ,
or in a layered volume of thickness $e_z$ (figure \ref{fi:milieu_spheres}).

\begin{figure}
\epsfxsize=5cm
\centerline{\epsfbox{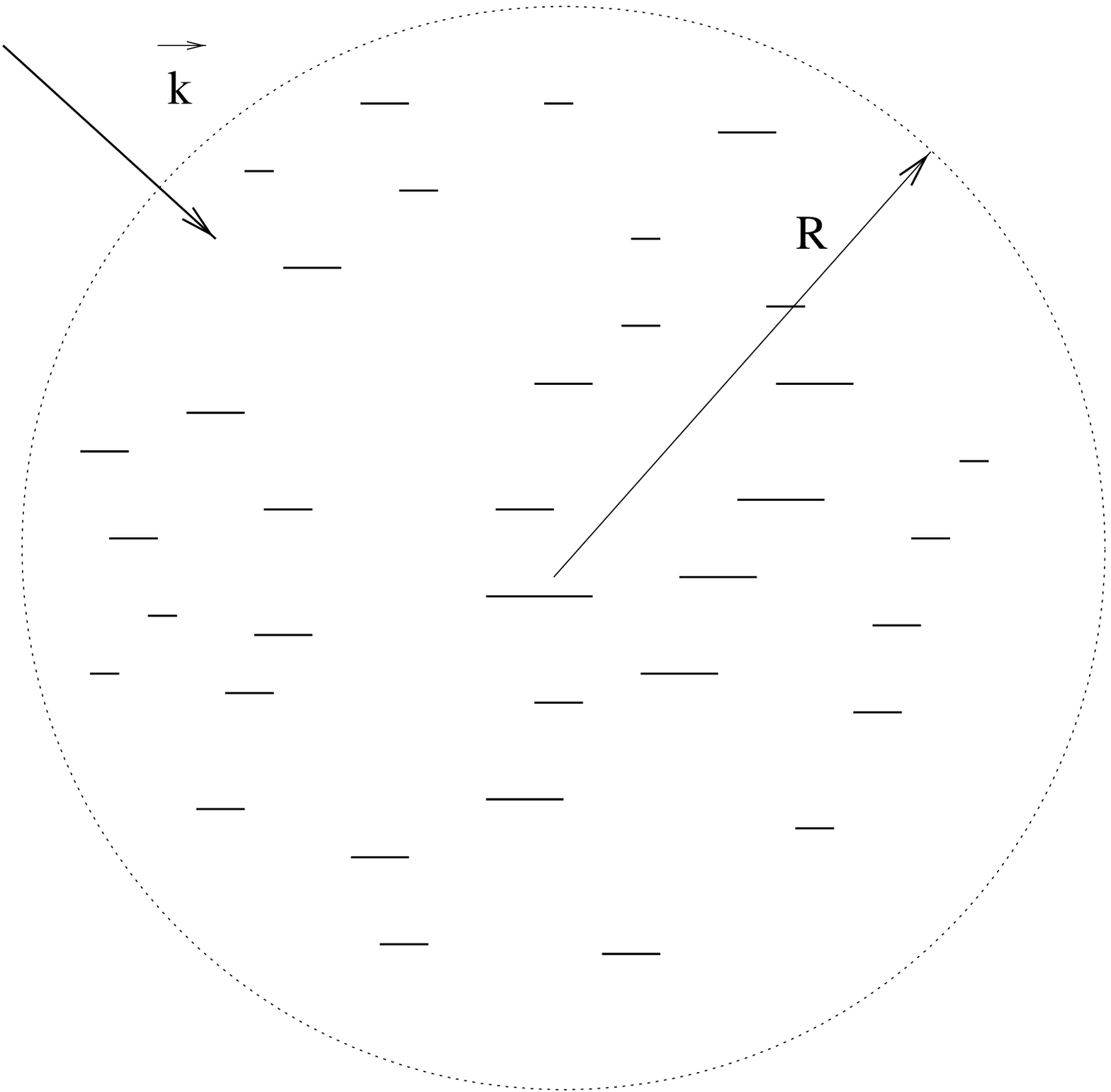}
\hskip 0.5cm 
\epsfxsize=5cm
\epsfbox{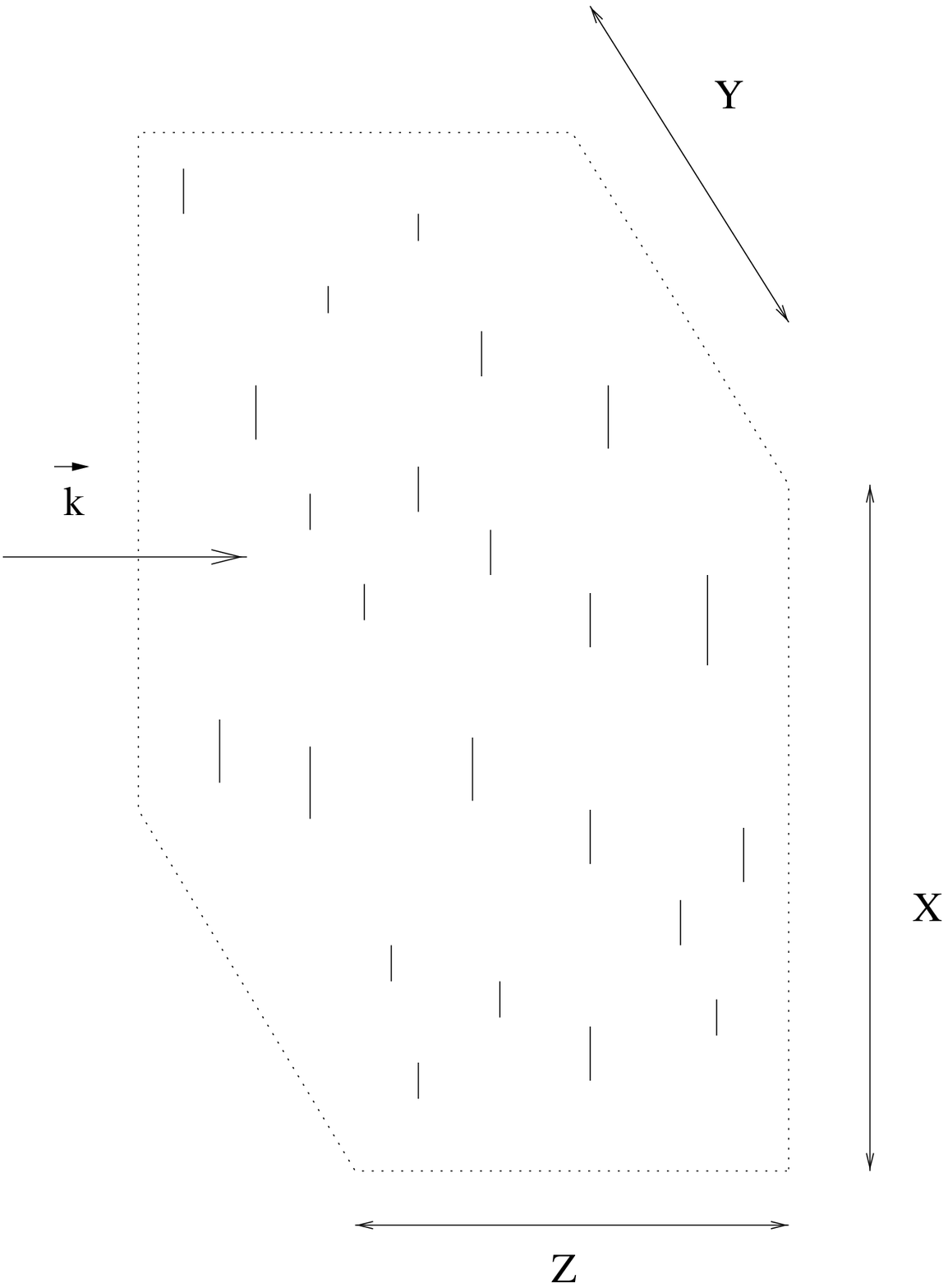}}
\vskip 2cm
\caption{Scheme for $N$ disks uniformly distributed in a spherical
volume or in a layered volume of finite thickness.}
\label{fi:milieu_disques}
\end{figure}

- A set of $N$ identical perfectly conducting disks of radius $a$ and
relative permittivity $\epsilon_{s}$, uniformly distributed in a spherical
volume $V$ , or in a layered volume of thickness $e_z$ (figure
\ref{fi:milieu_disques}).

\noindent
The definitions of the averaged wavevector $k_{e}$ and the scattering length
$l$ are given by the relations (\ref{ke}) and (\ref{l}). To express the
$k_{e}.l$ product, we have to calculate the self-energy function for
every disordered medium.

\subsection{Macroscopic homogenization conditions}

\noindent
First, we need to verify some conditions on the media in order to consider 
them as macroscopically homogeneous, which means that the self-energy function
must be only $k$ dependent.

The self-energy function in Fourier space for a medium  of $N$ identical
scatterers of scattering function $t(\overrightarrow{k},\overrightarrow{k}')$ 
is given by
\begin{align}
\Sigma(\overrightarrow{k},\overrightarrow{k}') = \frac{N}{V}
t(\overrightarrow{k},\overrightarrow{k}')
\int_{V}e^{i(\overrightarrow{k}-\overrightarrow{k}').
\overrightarrow{R}_{i}}d\overrightarrow{R}_{i},
\label{vdelta}
\end{align}
where $V$ is the volume of the medium. The integral in (\ref{vdelta}) taken
over a finite volume can be approximated as
\begin{align}
\label{homo}
\int_{V}e^{i(\overrightarrow{k}-\overrightarrow{k}').\overrightarrow{R}_{i}}
d\overrightarrow{R}_{i} =\delta(\overrightarrow{k}-\overrightarrow{k}') \quad \text{if}
\begin{cases}
\text{radius $R$ of the medium $>> \displaystyle{\frac{\lambda}{4}}$}\\
\text{thickness $e_z >> \displaystyle{\frac{\lambda}{2}}$}
\end{cases}
\end{align}
which leads to the condition of macroscopic homogenization, because the delta
distribution argument $\overrightarrow{k}'-\overrightarrow{k}$ implies that 
the self-energy function is only $\overrightarrow{k}$ dependent.

\subsection{Calculation of $k_{e}$ and $l$}

\noindent
Using expansions (\ref{devsigma}) and (\ref{multidif}), the first and the
second scattering order of the self energy operator are respectively given by
the following expressions
\begin{align}
\Sigma = \EE\big(\sum_{i=1}^{N} t_{i}\big),
\end{align}
\begin{align}
\Sigma = \EE\big(\sum_{i=1}^{N} t_{i}\big) - \EE\big(\sum_{i=1}^{N}
t_{i}.{\mathcal G}^{0}.t_{i}\big).
\end{align}
By assuming the macroscopic homogenization hypothesis, we can derive the
scalar expressions of the kernels.  For the second scattering order, 
we suppose that every scatterer receives a wave being locally plane,
in that case, the scalar part of the Green propagator simplifies to
$g^{0}(\overrightarrow{k},\omega) = -i
\pi \delta(k^2-k_{0}^2)$. Since the $N$ scatterers are uniformly distributed 
in the volume $V$, we obtain
\begin{align}
\label{simple}
\Sigma(\omega) = \frac{N}{V} t(\omega) \quad \text{single scattering order},
\end{align}
\begin{align}
\label{double}
\Sigma(\omega) = \frac{N}{V} t(\omega) + \pi \frac{N^2}{V^2} t^2(\omega) \quad
\text{second scattering order},
\end{align}
\noindent
where
\begin{align}
V = \begin{cases}
    \frac{4}{3}\pi R^3 \quad \text{for a spherical medium of radius $R$}\\
    e_x.e_y.e_z 
    \quad \text{for a layered volume of thickness $e_z$} \label{vol}\\
    \text{and transverse widths $e_x$ and $e_y$}
    \end{cases}
\end{align}
$R$ or $e_x$, $e_y$ and $e_z$ must satisfy the conditions (\ref{homo}). The
scattering functions are given by
\begin{align}
\label{tfonctions}
t(\omega) = \begin{cases}
              \frac{2i\pi}{k_{0}}\sum_{p=1}^{+\infty} (2p+1)(a_{p}+b_{p}) 
 \\ \label{sphere}
 \text{for a sphere of radius $a$ and permittivity $\epsilon_{s}$
\cite{Hulst}}\\
           \label{disque}
           -k_{0}^2 d \sqrt{2\pi}\bigg(\frac{16}{3}\big(\frac{k_{0}a}{2}\big)^2
+ \frac{512}{45}\big(\frac{k_{0} a}{2}\big)^4 + i
\frac{1024}{27\pi}\big(\frac{k_{0}a}{2}\big)^5\bigg) \\
\text{for a little disk of radius $a$ (d =2a) \cite{Nomura}} .
\end{cases}
\end{align}
\noindent
{}From (\ref{double}), (\ref{vol}), (\ref{disque}), we obtain the expression
of the averaged wavevector $k_{e}$ and the scattering length $l$ at the second
scattering order.

\vskip 0.5truecm
\noindent
- For $N$ spheres uniformly distibuted in a volume $V$
\begin{multline}
k_{e} = \frac{1}{\sqrt{2}}\sqrt{k^2 +
\frac{n}{2}\sum_{p=1}^{\infty}(2p+1)\mbox{Re}(a_{p}+b_{p}) + \frac{n^2}{4}
\big(\mbox{Re}(\sum_{p=1}^{\infty}(2p+1)(a_{p}+b_{p}))\big)^2
\big)} \\
\bigg[1 + \sqrt{1 +
\big(\frac{\frac{n}{2}\sum_{p=1}^{\infty}(2p+1)\mbox{Im}(a_{p}+b_{p})+
\frac{n^2}{4}\big(\mbox{Im}(\sum_{p=1}^{+\infty}(2p+1)(a_{p}+b_{p})\big))^2}
{k^2 +\frac{n}{2}\sum_{p=1}^{\infty}(2p+1)\mbox{Re}(a_{p}+b_{p})+\frac{n^2}{4}
\big(\mbox{Im}(\sum_{p=1}^{+\infty}(2p+1)(a_{p}+b_{p}))\big)^2}\big)^2}
\bigg]^{\frac{1}{2}},
\end{multline}
\noindent
and
\begin{align}
l =
\frac{k_{e}}{\frac{n}{2}\sum_{p=1}^{\infty}(2p+1)\mbox{Im}(a_{p}+b_{p})+
\frac{n^2}{4}\big(\mbox{Im}(\sum_{p=1}^{+\infty}(2p=1)(a_{p}+b_{p}))\big)^2}.
\end{align}
\vskip 1truecm
\noindent
- For $N$ disks uniformly distributed in a volume $V$
\begin{eqnarray}
k_{e} &=& \frac{1}{\sqrt{2}}\sqrt{k_{0}^2  -2n\sqrt{2\pi}k_{0}^2 d
\mbox{Re}(D_{0}^{0})+8\pi n^2 d^2 k_{0}^4(\mbox{Re}(D_{0}^{0})^2} \\ & &
\bigg[1 + \sqrt{1 + \big(\frac{-2n\sqrt{2\pi}k_{0}^2 d \mbox{Im}(D_{0}^{0})
+8\pi n^2
d^2 k_{0}^4\big(\mbox{Im}(D_{0}^{0})\big)^2}{k_{0}^2 -2n\sqrt{2\pi}k_{0}^2 d
\mbox{Re}(D_{0}^{0})+8\pi n^2 d^2 k_{0}^4
\big((\mbox{Re}(D_{0}^{0})\big)^2}\big)^2}\bigg]^{\frac{1}{2}},
\end{eqnarray}
\noindent
and
\begin{align}
l=\frac{k_{e}}{-2n\sqrt{2\pi}k^2 d \mbox{Im}(D_{0}^{0}) + 8\pi n^2 d^2 k_{0}^4
\big(\mbox{Im}(D_{0}^{0})\big)^2},
\end{align}
\noindent
where $D_{0}^{0} = \frac{16}{3}\big(\frac{k_{0}a}{2}\big)^2 +
\frac{512}{45}\big(\frac{k_{0} a}{2}\big)^4 + i
\frac{1024}{27\pi}\big(\frac{k_{0}a}{2}\big)^5$,
and $n = N/V$ is the density.

\section{Numerical results}

\subsection{Characterization of the localization domains}

We are now in position to search for localization domains, i.e, ranges of 
wavelength satisfying the localization criterion 
$k_{e}.l \leq \sqrt{\displaystyle\frac{3}{\pi}}$. In the following
we have computed the $k_{e}.l$ product as a function of the incident wavelength
$\lambda$, and labeled the localization curve as $k_{e}.l(\lambda)$. 
Then we will look for the influence of the system parameters on the 
localization, for instance: the number $N$, the width $a$ of the
scatterers and for the case of spheres, their permittivity $\epsilon_{s}$, the
volume $V$ of the medium. In our simulations, we have supposed that the
relative permittivity $\epsilon_{s}$ is weakly dependent on the pulsation, 
so we have treated it as a scalar. 
However, it is possible to give a description by a frequency law dependent 
upon the nature of the scatterer. 
The calculations were performed using {\it Mathematica} .

\begin{figure}
\epsfxsize=10cm
\centerline{\epsfbox{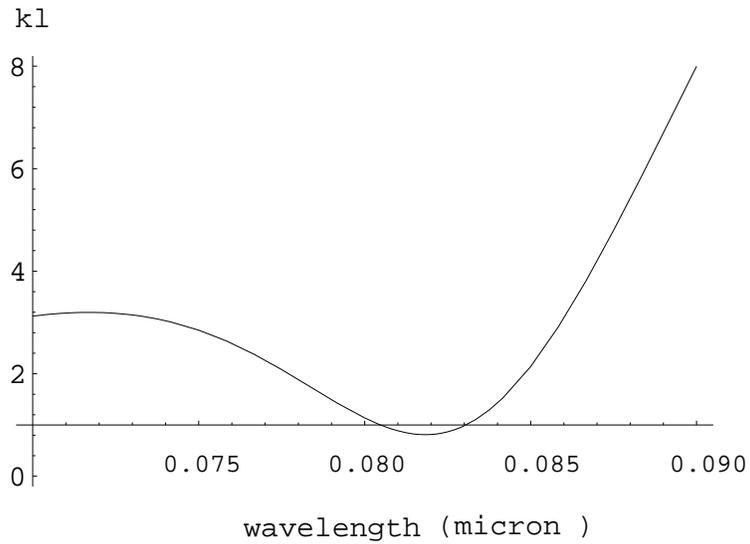}}
\vskip 1cm
\caption{Localization curve for a set $N=$ 1 million spheres of radius 
$a=0.01\mu$, and relative permittivity $\epsilon=16$, uniformly distributed 
in a finite spherical volume of radius $R=2 \mu$.}
\label{loca1}
\end{figure}

\vskip 0.5truecm
The first medium we study is an array of $N$ dielectric spheres of radius $a$,
permittivity $\epsilon_{s}$, uniformly distributed in a volume $V$ of radius
$R$. We numerically found a localization domain for the following
set of parameters (figure \ref{loca1})

\vskip 0.1truecm
\noindent
$N=$ 1 million\\
\noindent
a = 0.01$\mu$ \\
\noindent
$\epsilon_{s}$=16\\
\noindent
$R$ = 2$\mu$\\
\noindent
$\lambda \in [0.0805 \mu, 0.083 \mu]$.
\vskip 0.1truecm
\noindent
We satisfy the criteria of macroscopic homogenization of the medium
because in that case the ratio $\displaystyle{\frac{R}{\lambda}}$ 
is about 25, and so the condition
$\displaystyle{\frac{R}{\lambda}} >> \displaystyle{\frac{1}{4}}$ is valid. 
Let us notice that we are working in the range $k.a < 1$, and
only the first term of the expansion (\ref{tfonctions}) was used in the 
calcultations of the scattering function for the spheres.

\noindent
Now, we will investigate the effect of the parameters on the localization
domain.

\begin{figure}
\epsfxsize=10cm
\centerline{\epsfbox{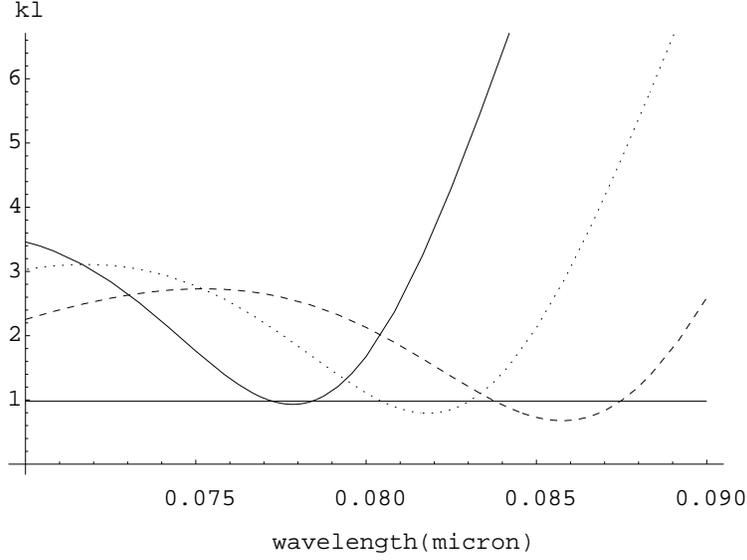}}
\vskip 1cm
\caption{Localization curves for a set of $N=$ 1 million spheres, of relative
permittivity $\epsilon_{s}=16$, and respective radii $a=0.095$ (plain curve),
$a=0.01$ (dotted) and $a=0.0105\mu$  (dashed), uniformly 
distributed in a finite spherical volume of radius $R=2 \mu$.}
\label{fi:rayon_sph}
\end{figure}

We show in figure \ref{fi:rayon_sph}, the localization curves for several 
values of the spheres radius. The dotted curve is taken for reference 
($a=0.01 \mu$). When the value of $a$ decreases (0.095$\mu$)
(plain curve), the minimum of the curve
occurs for lower wavelength, the corresponding value of the $k_{e}.l$ product
increases and the localization domain narrows. Oppositely, when the
radius is increased (0.0105$\mu$)(dashed curve), the minimum of the 
localization curve occurs for higher wavelength, the corresponding value 
of the $k_{e}.l$ product decreases and the localization domain enlarges.

\begin{figure}
\epsfxsize=10cm
\centerline{\epsfbox{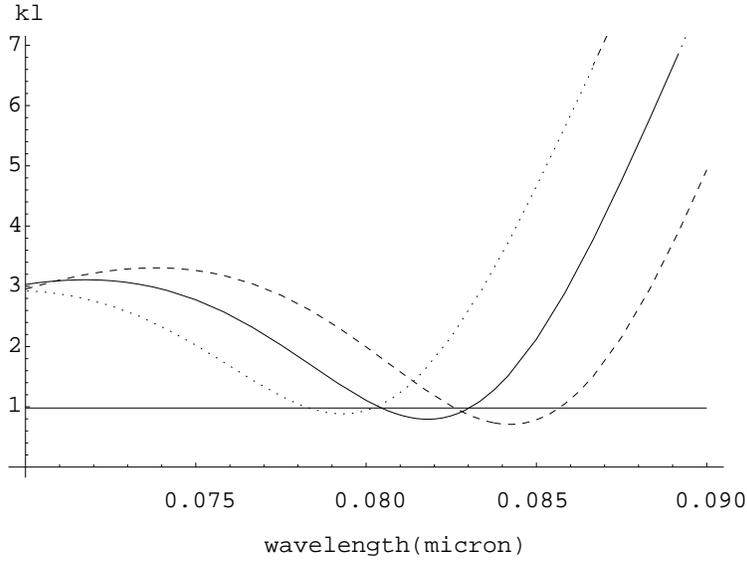}}
\vskip 1cm
\caption{Localization curves for a set of $N=$ 1 million spheres, of radius
$a=0.01 \mu$, and respective relative permittivities $\epsilon_{s}=15$ (plain
curve), $\epsilon_{s}=16$ (dotted) and $\epsilon_{s}=17$ (dashed),
uniformly distributed in a finite spherical volume of radius $R=2 \mu$.}
\label{fi:permit}
\end{figure}

In figure \ref{fi:permit}, we start from the reference configuration, and we
modify the relative permittivity of spheres. The curves are shown for the
values: $\epsilon=15$ (plain curve), $\epsilon=16$ (dotted curve) and
$\epsilon=17$ (dashed curve). When we slightly increase the relative 
permittivity of the spheres, the localization domain enlarges 
with a translation to higher wavelengths.

\begin{figure}
\epsfxsize=10cm
\centerline{\epsfbox{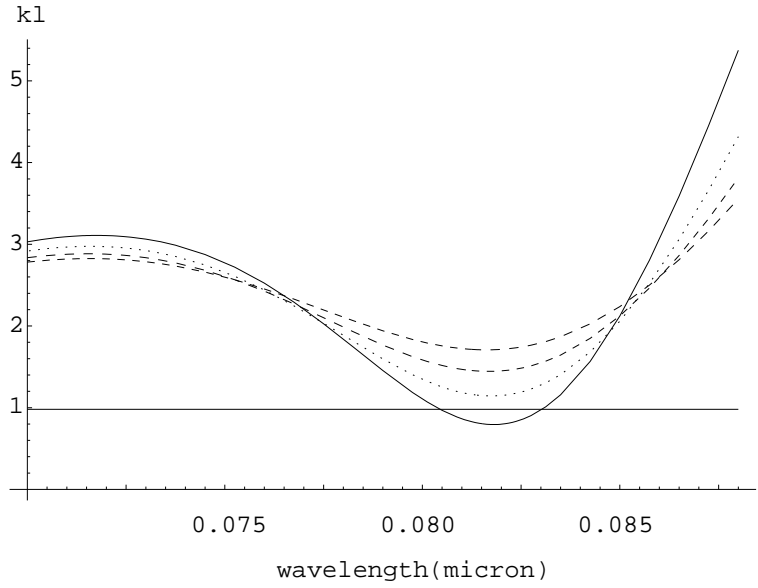}}
\vskip 1cm
\caption{Localization curves for a set of $N=$ 1 million spheres, of radius
$a=0.01 \mu$, with a relative real permittivity $\epsilon_{s}=16$,
and respective relative complex permittivities:
0 (plain curve), 0.4 (dotted), and 0.8 (dashed) and 1.2 (long dashed), 
uniformly distributed in a finite spherical volume of radius $R=2 \mu$.}
\label{fi:imag}
\end{figure}

Next, we have fixed the real part of the permittivity 
($\mbox{Re}(\epsilon)=16$), and we simulate a absorption on the surface 
of spheres by adding an imaginary part to the relative permittivity, 
so we obtain a dissipative dielectric. 
The figure \ref{fi:imag} shows the
localization curves for the respective values 0 (plain curve), 0.4 (dotted
curve), 0.8 (dashed curve) and 1.2 (long dashed curve) of the imaginary part of
the relative permittivity. When the value of the imaginary part increases,
the value of the minimum of the localization curve also increases and 
the localization domain narrows and finally disappear. 
By adding an imaginary part to the relative permittivity, the scatterers absorb 
a part of the electromagnetic energy and the interactions during successive 
scatterings are reduced, which explains the disappearance of localization.

An other interesting parameter is the density $n=\displaystyle{\frac{N}{V}}$. 
Indeed, when we
increase the density of scatterers, we also increase the backscattering
contributions from the successive interactions of the wave with scatterers, 
as a consequence the localization is more easily attained.

\begin{figure}
\epsfxsize=10cm
\centerline{\epsfbox{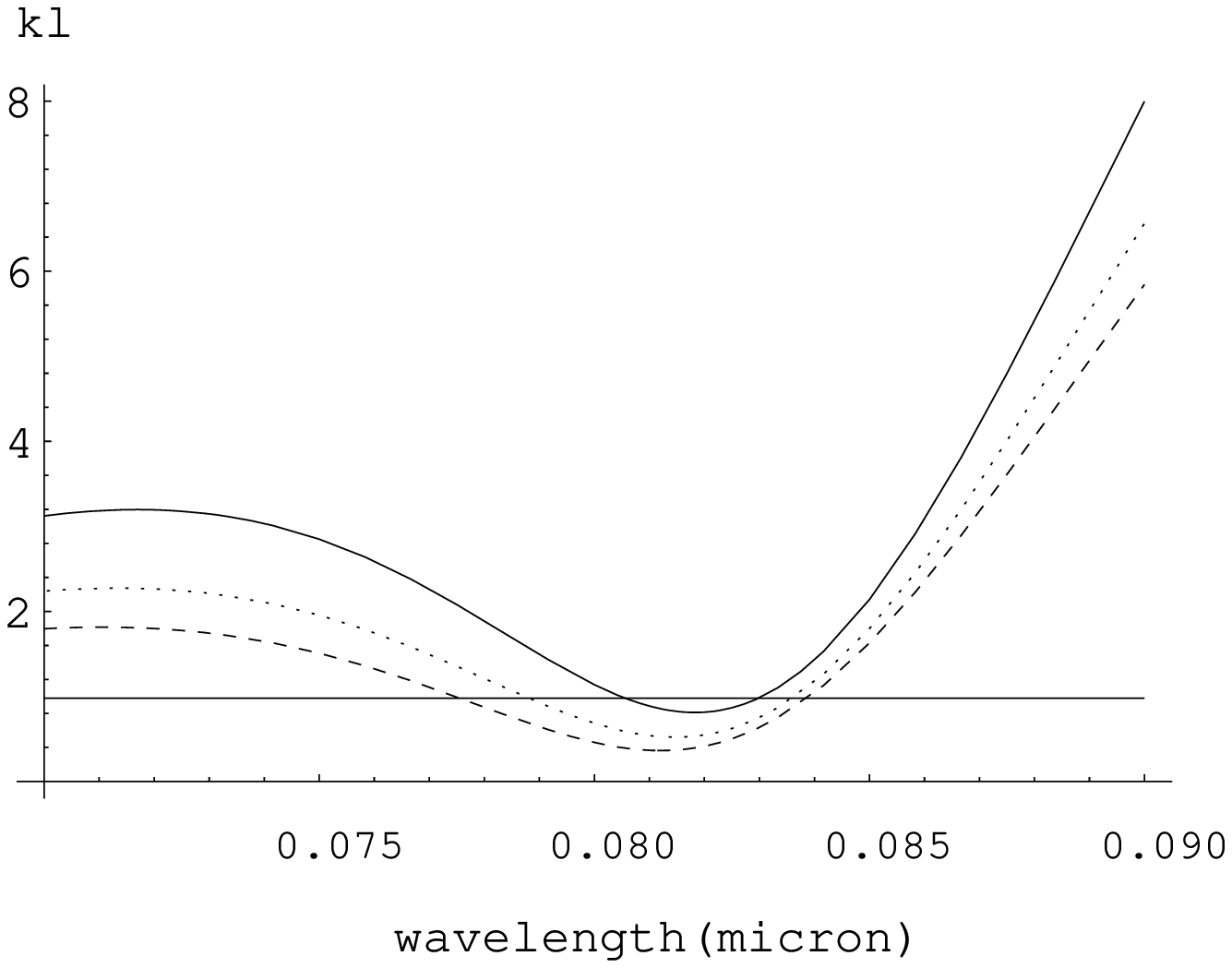}}
\vskip 1cm
\caption{Localization curves for the respectives sets of $N=1$ million (plain
curve), $N=1.5$ millions (dotted) and $N=2$ millions (dashed) spheres, of
radius $a=0.01 \mu$, and relative permittivity $\epsilon_{s}=16$, uniformly
distributed in a finite spherical volume of radius $R=2 \mu$.}
\label{fi:nombre_sph1}
\end{figure}

\begin{figure}
\epsfxsize=10cm
\centerline{\epsfbox{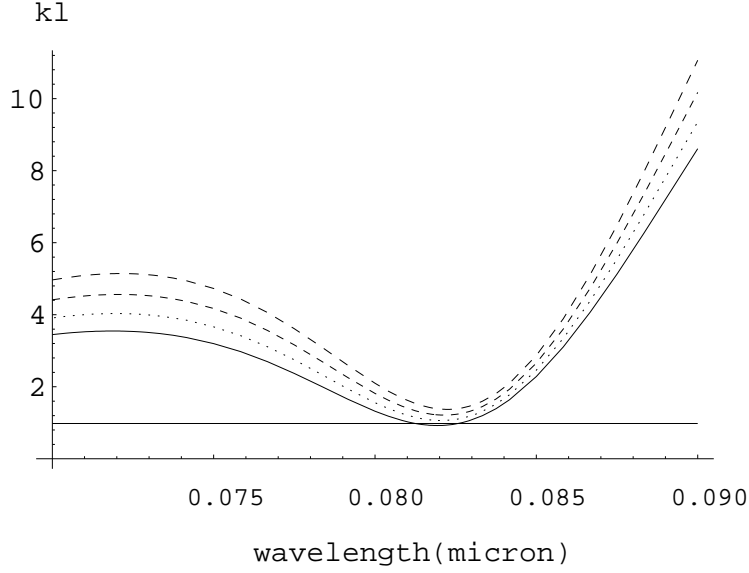}}
\vskip 1cm
\caption{Localization curves of a set of $N=$ 1 million spheres of radius
$a=0.01 \mu$, and relative permittivity $\epsilon=16$, uniformly distributed
in finite volumes of respective radii $R =$ 2.1 (plain curve), 2.2 (dotted),
2.3 (dash) and 2.4 $\mu$ (long dashed).}
\label{fi:rayon_milieu}
\end{figure}

We verify the influence upon the number $N$ of scatterers in a fixed volume
medium. The Figure \ref{fi:nombre_sph1} shows the localization curves
corresponding to $N=1$ million (plain curve), $N=$1.5 millions (dotted curve) 
and $N=$2 millions (dash curve) spheres.
When the number of scatterers increases, we observe a deepening of the
localization curve minimum and a broadening of the localization domain.
The inverse phenomenon was obtained when we fix the number of spheres and 
at the same time increase the radius of the medium. 
Figure \ref{fi:rayon_milieu} shows the localization curve for the radius $R$ 
values 2, 2.1, 2.2, 2.3, and 2.4 $\mu$ respectively. 
In that case, the minimum of the localization curve increases with $R$
and the localization domain disappears. 
The localization gradually vanishes because the medium is diluted, so the 
interactions between the scatterers are reduced. 
In both cases, the $N$ or $R$ variations show that the location of the minimums 
are weakly dependent, it means that one can choose a configuration giving
a central vawelength localization, and then modify the width of the 
localization domain around this value by the mean of medium density.

\vskip 0.1truecm
In the next part, we only focuse on the density effect.

\vskip0.1truecm
In the following example, we consider $N$ dielectric spheres uniformly
distributed in a rectangular volume 
$[-\frac{e_x}{2},\frac{e_x}{2}]\times
[-\frac{e_y}{2},\frac{e_y}{2}]\times [-\frac{e_z}{2},\frac{e_z}{2}]$, 
where $e_x$ and $e_y$ are large in front of $\lambda$ in order to 
approximate the rectangular medium by a single layer of thickness $e_z$.

\begin{figure}
\epsfxsize=10cm
\centerline{\epsfbox{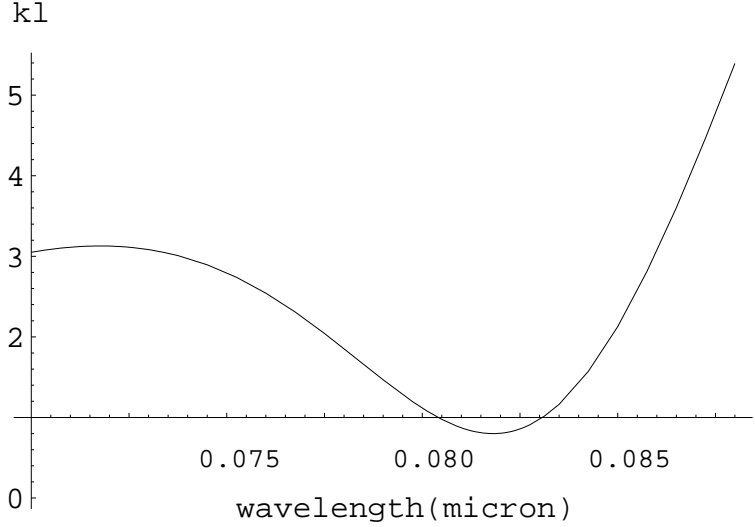}}
\vskip 1cm
\caption{Localization curve of a set of $N=10^5$ spheres of relative
permittivity $\epsilon_{s}=16$, and radius $a=0.01 \mu$ uniformly distributed 
in a layered volume of thickness $e_z=1.5 \mu$, and transverse lengths 
$e_x=1.5 \mu$ and $e_y=1.5 \mu$.}
\label{loca2}
\end{figure}

We numerically found a localization domain for the set of parameters (figure
\ref{loca2})

\vskip 0.1truecm
\noindent
$N=10^5$\\
\noindent
a = 0.01$\mu$ \\
\noindent
$\epsilon_{s}$=16\\
\noindent
thickness $e_z$ = 1.5 $\mu$\\
\noindent
transverse lengths $e_x$=$e_y$=1.5 $\mu$\\
\noindent
$\lambda \in [0.0805 \mu, 0.083 \mu]$.

\begin{figure}
\epsfxsize=10cm
\centerline{\epsfbox{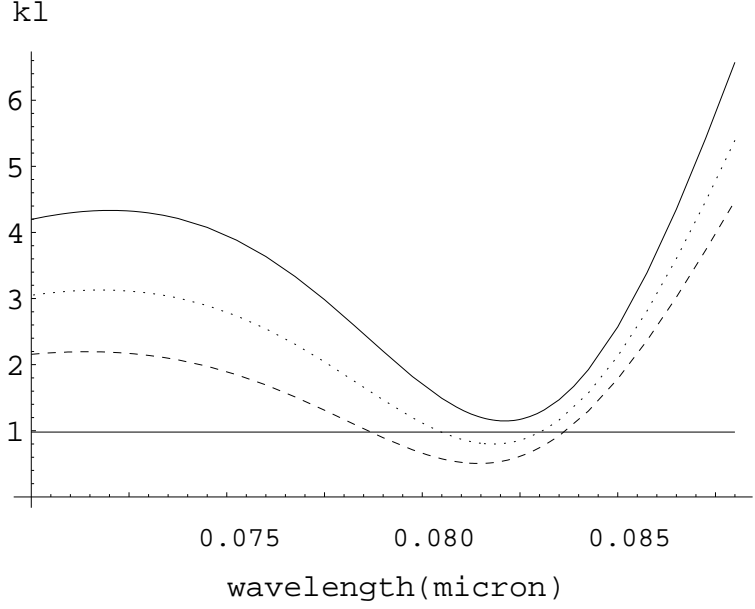}}
\vskip 0.5cm
\caption{Localization curves for the set of $N=10^5$ (plain curve), 
$N =1.2 10^5$ (dotted) and $1.5 10^5$ (dashed) spheres of radius  $a=0.01 \mu$,
and relative permittivity $\epsilon_{s}=16$, uniformly distributed
in a medium of thickness $e_z=1.5 \mu$ and of transverse lengths
$e_x=1.5 \mu$ and $e_y=1.5 \mu$.}
\label{fi:sph_couche_nombre}
\end{figure}

\begin{figure}
\epsfxsize=10cm
\centerline{\epsfbox{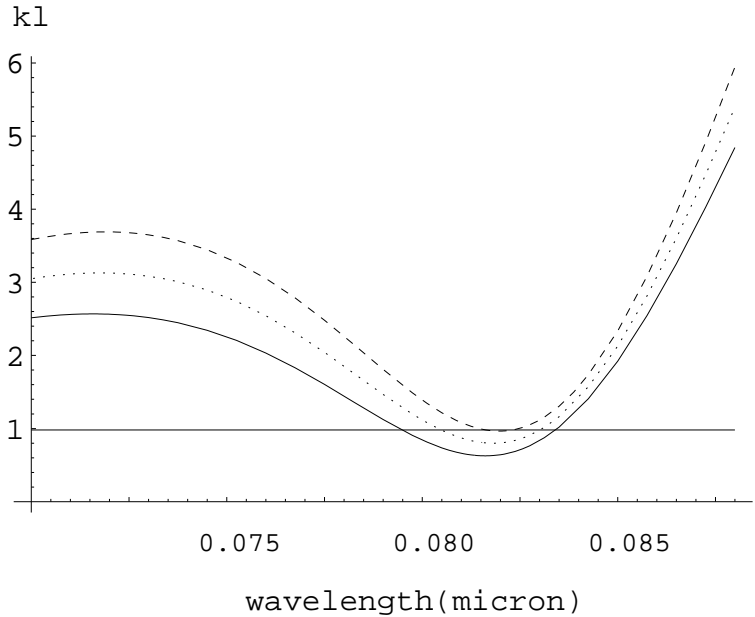}}
\vskip 0.5cm
\caption{Localization curves for a set of $N=10^5$ spheres of radius $a=0.01
\mu$ and relative permittivity $\epsilon_{s}=16$, uniformly distributed in
different media of respective thickness $e_z=1\mu$ (plain curve), 
$e_z=1.5\mu$ (dotted) and $e_z=1.8\mu$ (dashed), and of respective lengths
$e_x=1.5\mu$ and $y=1.5 \mu$.}
\label{fi:sph_couche_ep}
\end{figure}

\vskip 0.1truecm
The figure \ref{fi:sph_couche_nombre} presents the localization curves when 
the number $N$ of spheres takes respectively the values:
$10^5$ (plain curve), $1.2 10^5$ (dotted curve) and $1.5 10^5$ (dashed curve),
the other parameters are kept fixed. 
We observe that the minimum of the curves diminishes with the increase of $N$
and the localization domain enlarges. Then, if we fix the number
of spheres $N = 10^5$, and modify the thickness $e_z$ of the layer with respect
to the macroscopic homogenization condition. The figure \ref{fi:sph_couche_ep}
shows the curves obtained for $e_z= 1\mu$ (plain curve), $e_z= 1.5\mu$ 
(dotted curve) and $e_z= 1.8 \mu$ (dashed curve). 
The minimum of the curves increases with the thickness $e_z$ of the layer,
the localization domain narrows and finally disappears.
The observed effects are the same as in the case of a spherical volume.

\vskip 0.1truecm
In a second series of tests, we replace the spheres by perfectly conducting
disks, oriented following a plane perpendicular to the wavevector (figure
\ref{fi:milieu_disques}).

\begin{figure}
\epsfxsize=10cm
\centerline{\epsfbox{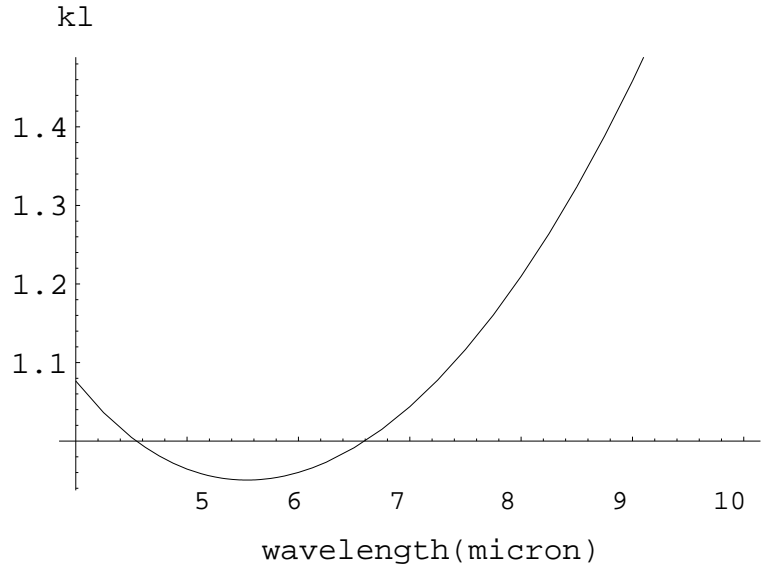}}
\vskip 1cm
\caption{Localization curve for a set of $N$ = 1 billion of perfectly
conducting disks of radius $a=0.05 \mu$, uniformly distributed in spherical
volume of radius $R=12.5 \mu$.}
\label{loca3}
\end{figure}

For the two respective media (spherical and layered),
we numerically find a localization domain for the set of parameters
\underline{in a spherical volume} (figure \ref{loca3}),

\vskip 0.1truecm
\noindent
$N=1$  billion\\
\noindent
a = 0.05$\mu$ \\
\noindent
volume radius $R$ = 12.5$\mu$\\
\noindent
$\lambda \in [0.45 \mu, 0.64 \mu]$

\begin{figure}
\epsfxsize=10cm
\centerline{\epsfbox{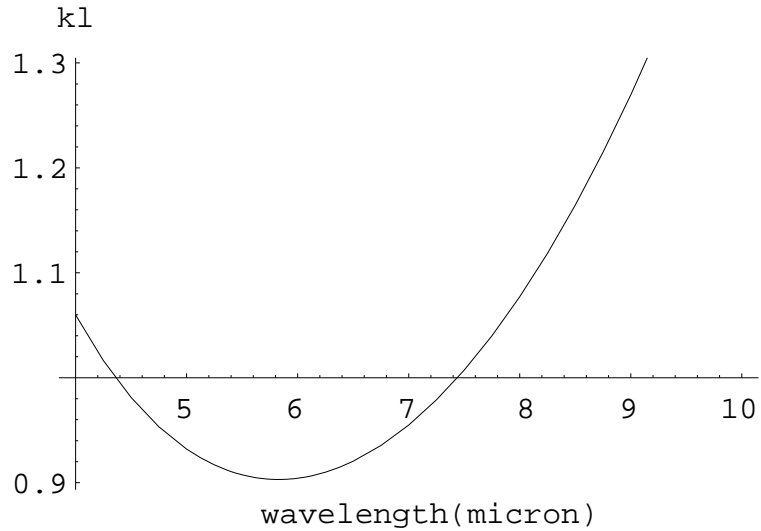}}
\vskip 1cm
\caption{Localization curve for a set of $N$=150 millions perfectly conducting
disks, of radius $a=0.05 \mu$ uniformly distributed in a layered volume of
thickness $e_z=10\mu$ and transverse widths $e_x=10 \mu$ and $e_y=10 \mu$.}
\label{fi:loca4}
\end{figure}

\noindent
\underline{and for a single layer} (figure \ref{fi:loca4})

\vskip 0.1truecm
\noindent
$N=150 $millions\\
\noindent
a = 0.05 $\mu$ \\
\noindent
thickness $e_z$ = 10 $\mu$\\
\noindent
transverse lengths $e_x$=$e_y$=10 $\mu$\\
\noindent
$\lambda \in [0.45 \mu, 0.74 \mu]$.

\begin{figure}
\epsfxsize=10cm
\centerline{\epsfbox{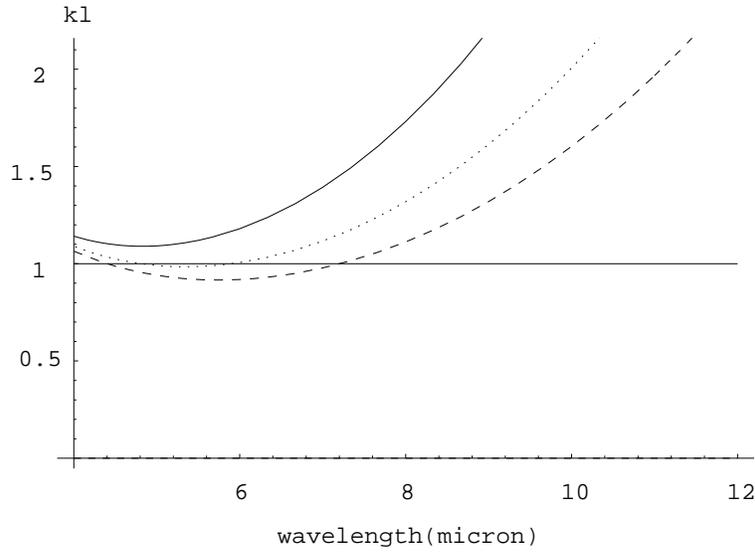}}
\vskip 1cm
\caption{Localization curves for the respective sets of $N$=1 (plain curve),
$N$=1.5 (dotted), $N$=2 billions (dashed) perfectly conducting disks
of radius $a=0.05\mu$, uniformly distributed in a spherical volume of radius
$R=15 \mu$.}
\label{fi:disque_nombre_sph}
\end{figure}

\begin{figure}
\epsfxsize=10cm
\centerline{\epsfbox{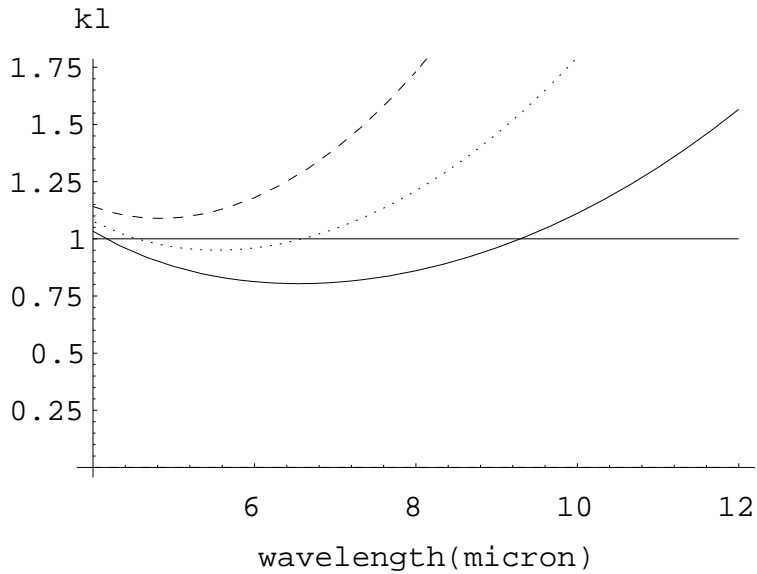}}
\vskip 1cm
\caption{Localization curves for a set of $N$=1 billion disks of radius
$a=0.05 \mu$ uniformly distributed in different media of respective radii
$R=10$ (plain curve), $R$=12.5 (dotted), $R=15 \mu$ (dashed).}
\label{fi:disque_volume_sph}
\end{figure}

\vskip 0.1truecm
Figures \ref{fi:disque_nombre_sph} and \ref{fi:disque_volume_sph} show the
localization curves for the spherical volume case when we vary respectively 
the number of disks and the radius of the volume. In figure
\ref{fi:disque_nombre_sph}, the volume radius is fixed, $R=15\mu$, and
the number $N$ takes the values 1 billion
(plain curve),  1.5 billions (dotted curve) and 2 billions (dashed curve), 
for this last value, the localization domain is located in the range [0.45
$\mu$, 0.72 $\mu$], it is larger compared to the case $N$=1.5 billion
([0.45 $\mu$, 0.60 $\mu$]).
In figure  \ref{fi:disque_volume_sph}, we have fixed $N$ at 1 billion and
computed localization curves for $R$ = 10 $\mu$ (plain curve), $R$=12.5
$\mu$ (dotted curve), and $R$ = 15 $\mu$ (dashed curve). In the last
case, the minimum of the localization curve is 1.1 for $\lambda=5 \mu$, but
there is no more localization.

\begin{figure}
\epsfxsize=230pt
\centerline{\epsfbox{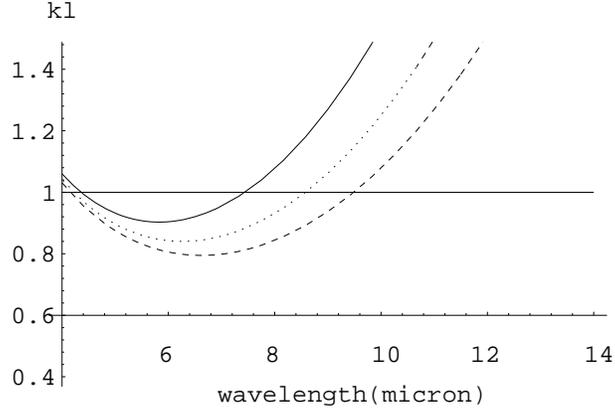}}
\vskip 1cm
\caption{Localization curves for the respective sets of $N$=150 (plain curve),
$N$=200 (dotted) and $N$=250 millions (dashed) perfectly conducting
disks of radius $a=0.05 \mu$, uniformly distributed in a layered medium of
thickness $e_z=10 \mu$, and transverse widths $e_x=10 \mu$ and $e_y=10 \mu$.}
\label{fi:disque_nombre_cou}
\end{figure}

\begin{figure}
\epsfxsize=10cm
\centerline{\epsfbox{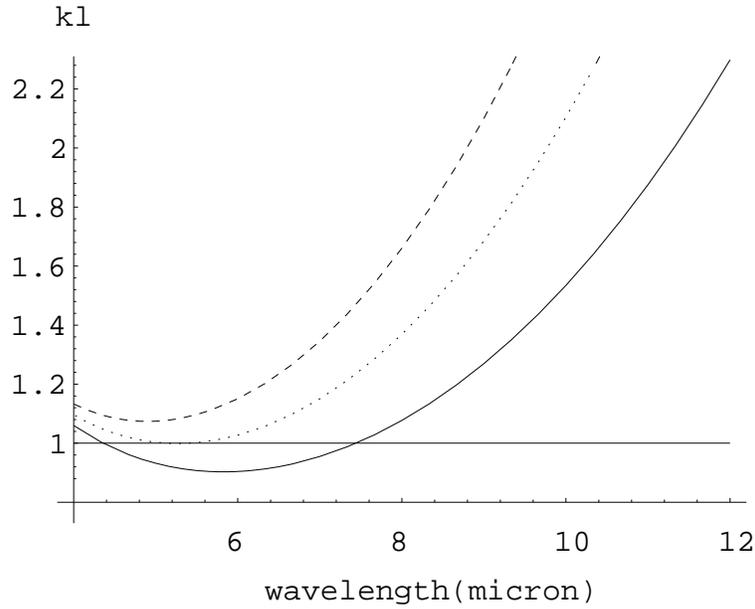}}
\vskip 1cm
\caption{Localization curves for a set of $N=150$ millions disks of radius
$a=0.05\mu$, uniformly distributed in several media of respective thickness 
$e_z=10\mu$ (plain curve), $e_z=15\mu$ (dotted) and $e_z=20\mu$ 
(dashed), and transverse widths $e_x=10\mu$ and $e_y=10\mu$.}
\label{fi:disque_volume_cou}
\end{figure}

The same effects occur in the case of a layered medium. When we increase
the number of disks (figure \ref{fi:disque_nombre_cou}) the localization 
domain enlarges, oppositely, when the thickness of the medium increases 
(figure \ref{fi:disque_volume_cou}), the localization range narrows 
and disappears.

\subsection{A Numerical check}

In order to numerically control the behaviour of the electromagnetic field near
a localization range, we have used a multidipolar diffusion formulation
\cite{tor1}. The principle consists to discretize a dielectric sphere of radius
$a < \lambda$ by an array of dipoles \cite{penny}, and to compute the
scattered field using a multidipolar expansion.  To characterize the scattered
field, we use the intensity functions $I_{ij}(\theta)$ described in
\cite{tor2}, where $i$ and $j$ respectively refer to the polarization of the
incident and scattered electric field, $\theta$ is the scattering
angle. More precisely, we are interested by the intensity behaviour 
around the backscattering direction, where the localization phenomenon
manifests by the creation of a backscattering peak.

We compute the averaged intensity functions by supposing the disordered medium
ergodic, which means that the mathematical expectation value of the intensity 
function is obtained by averaging intensity functions on a large number of 
disordered medium configurations.

Due to the large number of spheres we used for computations, we simplify the
problem by modeling a little sphere by a single dipole located at its center.

\subsection{Case of a spherical volume}

We first use the localization parameters previously obtained for the set of 
spheres in a spherical volume to compute the averaged intensity 
functions. In this model, a number of spheres around 1 million is too large
to be handled by numerical computations. But we notice that the fundamental
parameter which occurs in the localization criterion is the density 
$n=\displaystyle{\frac{N}{V}}$. To simulate an equivalent configuration, 
we define a medium of the same density but with a smaller number of spheres. 
However, the consequence is also to reduce the radius of the medium, 
and then alterate the macroscopic homogenization condition (\ref{homo}).

We choose the set of the following parameters: $R=0.2 \mu$ and $N=1000$. The
localization parameters are then

\vskip 0.1truecm
\noindent
$N=1000$\\
\noindent
a = 0.01 $\mu$ \\
\noindent
$\epsilon_{s}$=16\\
\noindent
$R$ = 0.2 $\mu$\\
\noindent
$\lambda \in [0.0805 \mu, 0.083 \mu]$

\begin{figure}
\epsfxsize=10cm
\centerline{\epsfbox{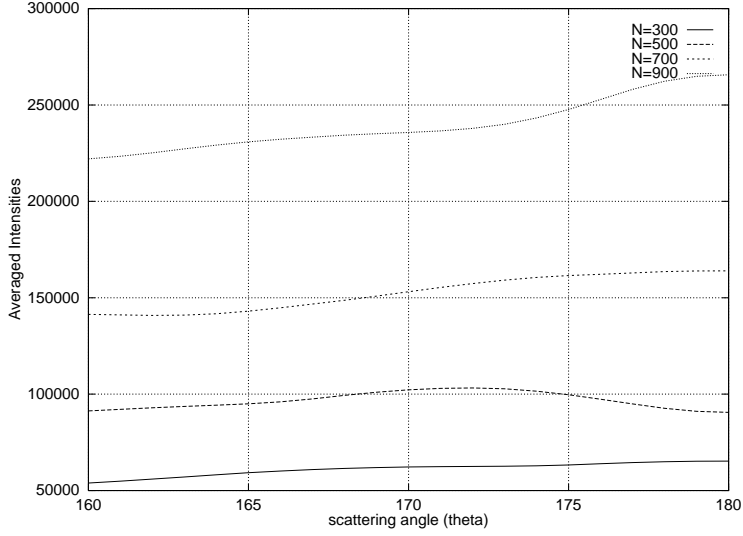}}
\vskip 1cm
\caption{Averaged intensities over 20 configurations as a function of the
scattering angle $\theta$, for an $x$ incident polarization of the 
electric field for different media, where $N$ =300, 500, 700, 900 spherical 
scatterers of radius $a=0.01\mu$, and relative permittivity $\epsilon_{s}=16$ 
uniformly distributed in a spherical volume of radius $R=2 \mu$. The incident 
wavelength is $\lambda_{0}= 0.082 \mu$.}
\label{fi:retro_sph}
\end{figure}

\vskip 0.1truecm
We propose to compute the intensity functions for several values of $N$, when
we approach a localization range (i.e., $N \rightarrow 1000$). Figure
\ref{fi:retro_sph} shows the intensity functions $I_{xx}(\theta)$ 
(in arbitray units) for the respective values of $N$ =300,500,700,900. 
We notice the appearance of a peak in the backscattering direction when the 
value of $N$ approaches the theoretical value of $N$ given by the mathematical model ($N$=1000).

\subsection{Case of a layered medium}

We use the localization parameters given for the set of spheres in a layered
volume to compute the averaged intensity functions. Like in the previous
example, the number of spheres is too large for a numerical computation.
To overcome this difficulty, we use an equivalent medium with the same density.

We choose the set of equivalent parameters: $e_x=0.3 \mu$, $e_y=0.3\mu$, 
$e_z=0.28\mu$ and $N=800$ . The localization parameters are then

\vskip 0.1truecm
\noindent
$N=800$\\
\noindent
$a=0.01 \mu$\\
\noindent
$\epsilon_{s}=16$\\
\noindent
$e_x=e_y=0.3 \mu$\\
\noindent
$e_z=0.28\mu$\\
\noindent
$\lambda \in [0.081 \mu,0.083 \mu]$

\begin{figure}
\epsfxsize=10cm
\centerline{\epsfbox{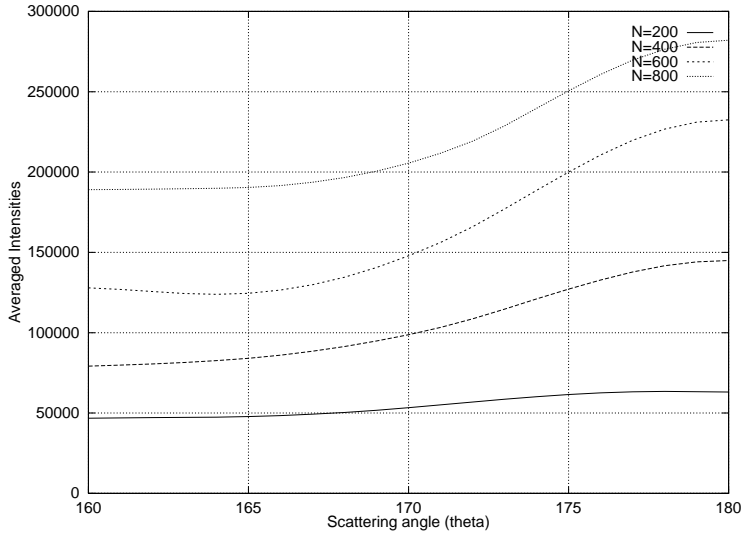}}
\vskip 1cm
\caption{Averaged intensities over 20 configurations as a function of the
scattering angle $\theta$ for an $x$ incident polarization of the electric 
field in a layered medium of thickness $z=0.28\mu$, where 
$N$=200, 400, 600, 800 spherical scatterers of radius $a=0.01 \mu$, 
and relative permittivity $\epsilon_{s}=16$. The incident wavelength is 
$\lambda_{0}=0.082 \mu$.}
\label{fi:retro_couche}
\end{figure}

\vskip 0.1truecm
We present the intensity function curves $I_{xx}(\theta)$ when the number of
spheres $N$ have respectively the values 200, 400, 600, 800 
(figure~\ref{fi:retro_couche}). We also observe a backscattering peak
when $N$ tends to the initial value where one observes an electromagnetic
energy localization in the mathematical model.

\section{Conclusion and perspectives}

We have analytically calculated the value of the $k_{e}.l$ product as 
a function of the incident wavelength in order to find wavelength 
domains satisfying the localization criterion 
$k_{e}.l \le \displaystyle{\sqrt{\frac{3}{\pi}}}$ in different media.
We have considered some configurations where the volume may be finite or
infinite, and the scatterers are dielectric spheres or perfectly conducting
little disks. We have studied the influence of electromagnetic system
parameters on the localization domain, for instance,
the complex permittivity of spheres, their radius, their number and the 
dimensions of the surrounding medium. Computations have shown that the density 
offers the optimal way to adjust the width of the localization domain.

The multidipolar model allowed us to detect a backscattering peak when the
system parameters are closed to the localization parameters provided with the
theoretical model. Many works were already realized in order to calculate the
expression of the backscattering peak \cite{Sheng}, \cite{Akkermann}, however
these results were not related to the localization criterion. This peak,
predicted by the theoretical model, occurs from the crossed diagrams
contribution \cite{wolfe} responsible of the backscattering, it represents a 
first {\it manifestation} of the localization phenomenon. 
However, we were limited by the size of the matrix describing the system, 
and the conditions where the medium is macroscopically homogeneous were not 
exactly fulfilled. 
The same matrix limitation implies that we have, in the simulation, described 
a little sphere by a single dipole. A way to perform a more realistic 
simulation would consist to describe the sphere itself by a set of dipoles
located on a cubic lattice inside the sphere \cite{penny}.

\section{Acknowledgements}

C. O. thanks Thomson-CSF Optronique for a financial support during the
preparation of his thesis. Contract CIFRE-400-95.

\setlength{\baselineskip}{15pt}
\newpage

\newpage

\newpage

\newpage

\newpage

\newpage

\newpage

\newpage

\newpage

\newpage

\end{document}